\documentclass{ws-procs975x65}
\usepackage{graphicx}
\usepackage{pifont}

\def\beq{\begin{equation}}
\def\eeq{\end{equation}}

\newcommand{\cmark}{\ding{51}}%
\newcommand{\xmark}{\ding{55}}%

\begin{document}
\title{BlackHoleCam $-$ Testing General Relativity with Pulsars Orbiting Sagittarius A*}
%\author{Andrew B. Author$^*$ and Charles D. Author}
\author{Ralph~P.~Eatough$^{*1}$, Gregory~Desvignes$^1$, Kuo~Liu$^1$, Robert~S.~Wharton$^1$, Aristedis~Noutsos$^1$, Pablo~Torne$^{2,1}$, Ramesh~Karuppusamy$^1$, Lijing~Shao$^{3,1}$, Michael~Kramer$^{1,4}$, Heino Falcke$^{5,1}$, Luciano Rezzolla$^{6}$}

\address{$^1$Max-Planck-Institut f{\"u}r Radioastronomie, Auf dem H{\"u}gel 69,\\
Bonn, D-53121, Germany\\
$^*$E-mail: reatough@mpifr-bonn.mpg.de\\
www.mpifr-bonn.mpg.de}

\address{$^2$Instituto de Radioastronom{\'i}a Milim{\'e}trica, IRAM, Avenida Divina Pastora 7,\\
 Local 20, E-18012, Granada, Spain\\}

\address{$^3$Kavli Institute for Astronomy and Astrophysics,\\
Peking University, Beijing 100871, China\\}

\address{$^4$Jodrell Bank Centre for Astrophysics, The University of Manchester,\\
 Alan Turing Building, Manchester M13 9PL, UK\\}

\address{$^5$Department of Astrophysics, Institute for Mathematics, Astrophysics and Particle Physics (IMAPP), Radboud University,P.O. Box 9010, 6500 GL Nijmegen, The Netherlands\\}

\address{$^6$Institut f{\"u}r Theoretische Physik, Goethe-Universit{\"a}t Frankfurt, Max-von-Laue-Stra{\ss}e 1,\\ D-60438 Frankfurt am Main, Germany\\}

%This article file typeset from main.tex gives a quick template to
%insert the correct titlepage material at the top of the standard
%more commonly known article.cls style formatting which is similar to
%many journal styles apart from titlepage details. The MG variation of
%the World Scientific proceedings macros removes the white space above
%the title, to give more text space for short articles.

\begin{abstract}
BlackHoleCam is a project funded by a European Research Council
\emph{Synergy Grant} to build a complete astrophysical description of
nearby supermassive black holes by using a combination of radio
imaging, pulsar observations, stellar astrometry and general
relativistic magneto-hydrodynamic
models. BlackHoleCam\footnote{www.blackholecam.org} scientists are
active partners of the Event Horizon Telescope
Consortium\footnote{www.eventhorizontelescope.org}. In this talk I
will discuss the use of pulsars orbiting Sagittarius~A* for tests of
General Relativity, the current difficulties in detecting such
sources, recent results from the Galactic Centre magnetar
PSR~J1745$-$2900 and how BlackHoleCam aims to search for undiscovered
pulsars in the Galactic Centre.
\end{abstract}

%\keywords{Sample file; \LaTeX; MG15 Proceedings; World Scientific Publishing.}

\bodymatter

%%%%%%%%%%%%%%%%% now a standard article style for the most part

\section{The science case for observing pulsars orbiting Sagittarius~A*}
For over 40 years observations double neutron star systems $-$ the
collapsed and degenerate remnants of massive stars $-$ where one, or
both, stars are active radio pulsars have demonstrated that they form
exceptional natural ``laboratories'' for precision tests of theories
of gravitation\cite{tw82,ksm+06}. In this vein, a pulsar in a close
orbit around the supermassive black hole at the centre of our Galaxy
(Sagittarius~A* $-$ Sgr~A* for short) would be at the apex of
gravity experiments made possible using pulsars\cite{wk99,lwk+12}. 

\begin{figure}[h]
\begin{center}
\includegraphics[width=2.5in]{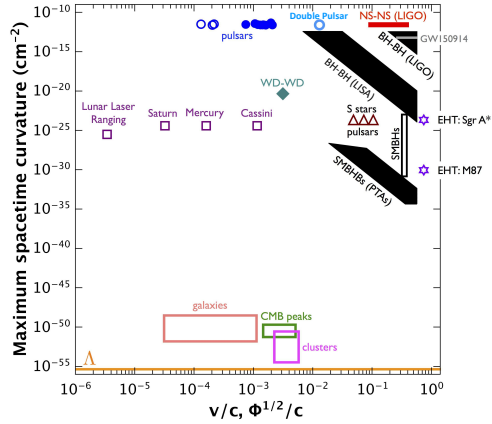}
\end{center}
\caption{Parameter space of observations and tests of gravity. On the
  x-axis, $v$ denotes the typical velocity of the system's components
  while $\Phi$ denotes the gravitational potential being probed by
  photons propagating in the corresponding spacetime. On the y-axis
  the maximum spacetime curvature (taken at the horizon for black
  holes) in the system is indicated as a measure of how much the
  system deviates from flat spacetime. Filled areas indicate
  gravitational wave tests, while hollow areas stand for
  quasi-stationary tests, including accretion onto compact
  objects. The rightmost hollow blue circle stands for the Shapiro
  delay test in the double pulsar. Figure and caption reproduced from
  Ref.~6 with the kind permission of Wex and Kramer.}
\label{fig1}
\end{figure}

\begin{figure}[h]
\begin{center}
\includegraphics[width=2.3in]{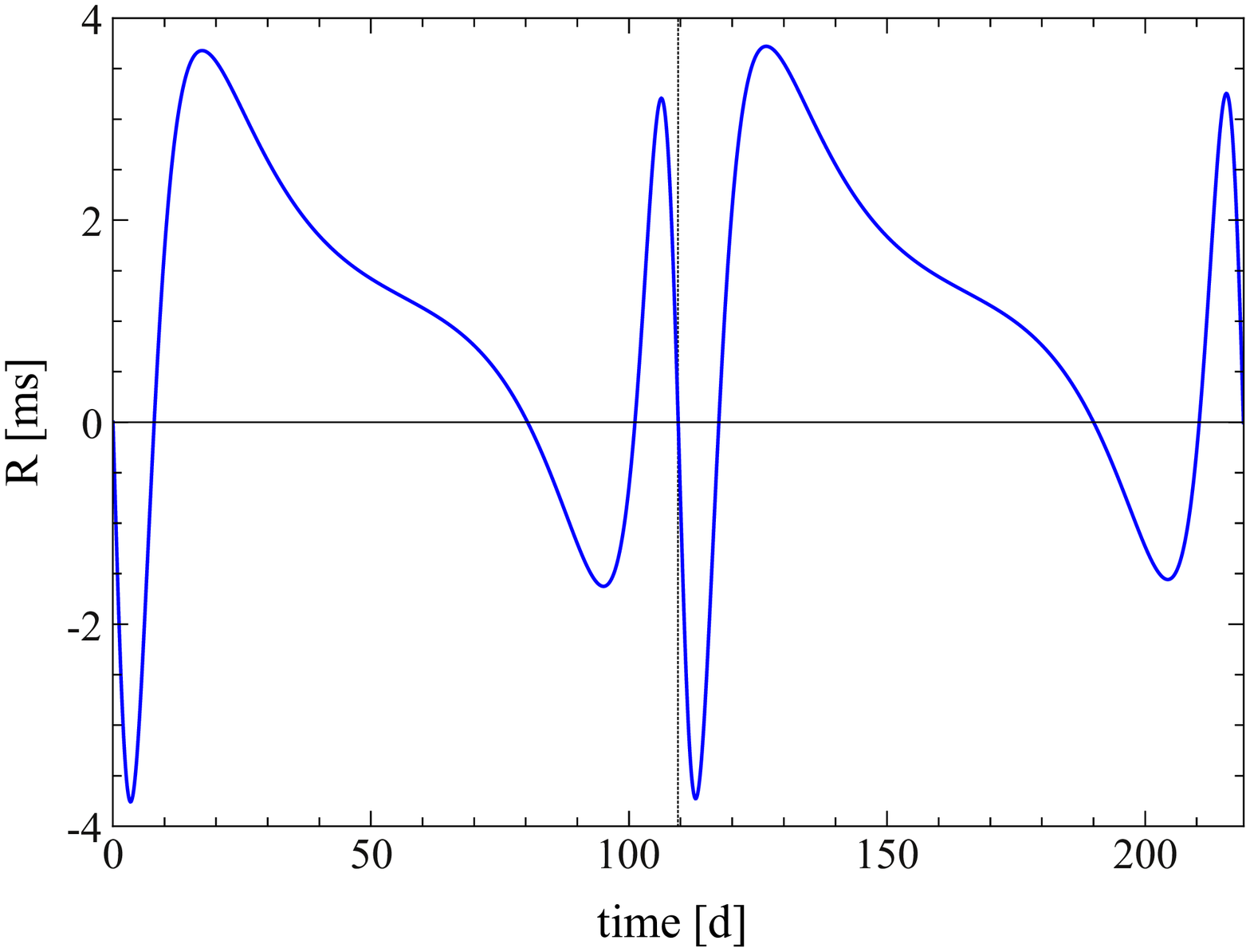}
\hspace*{5pt}
\includegraphics[width=1.93in]{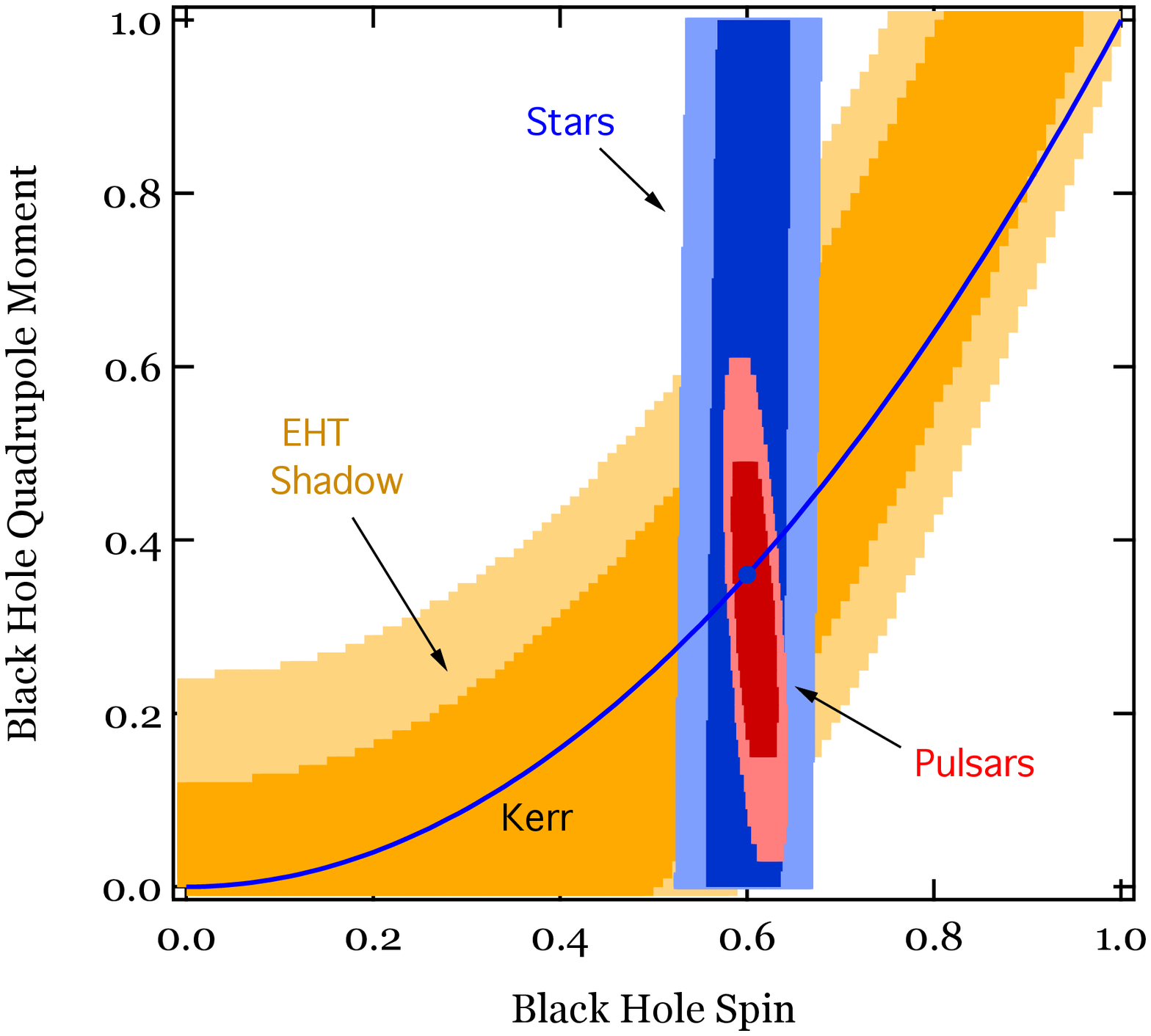}
\end{center}
\caption{(\emph{left}) A simulation showing the effect of the
  quadrupole moment of Sgr~A* ``imprinted'' upon pulsar timing
  residuals over two orbital cycles. Here an orbital period of
  0.3~yr, an eccentricity of 0.5 and a dimensionless black
  hole spin parameter of unity were assumed. Figure reproduced from
  Ref.~4. (\emph{right}) Comparison of the posterior
  likelihood of measuring the spin and quadrupole moment of Sgr~A*
  using the orbits of two stars (blue), timing of three periapsis
  passages of a normal pulsar (red) and the shape of the black hole
  shadow from radio interferometric imaging (gold).  The solid curve shows the
  expected relation between the two quantities for a Kerr metric. The
  filled circle marks the assumed spin and quadrupole moment. Figure
  reproduced from Ref.~7.}
\label{fig2}
\end{figure}

\noindent Such a system will allow the fundamental predictions of black-hole
properties in General Relativity (GR) to be tested; properties that
Advanced LIGO, which measures the strongly dynamical regime of a
merger, potentially cannot\cite{tll+17}. These include the {\em
  no-hair theorem} and the {\em cosmic censorship
  conjecture}\cite{wk99,lwk+12}. For example, the latter is tested
through measurements of frame dragging caused by the spinning black
hole, which manifests itself as a contribution to the precession of
the pulsar orbit. A Kerr black hole should exhibit a dimensionless
spin parameter $\chi$ which is no larger than unity; a spin parameter
greater would be in conflict with GR posing a direct
contest to the theory. Fig.~\ref{fig1} shows how pulsar tests fit
into the relativistic regime of other gravity tests past and
present. In Fig.~\ref{fig2}~({\emph left} panel) from
Ref.~\refcite{lwk+12} the expected signature of the black hole
quadrupole moment in pulsar timing residuals for a putative pulsar
closely orbiting Sgr~A* is displayed. Ref.~\refcite{pwk+16} have shown
that by combining measurements of the black hole spin from stellar
astrometry, pulsar timing and interferometric imaging of the black
hole shadow, an unbiased and quantitative test of the no-hair theorem is
possible (Fig.~\ref{fig2} \emph{right} panel).

%\subsection{Subsections only have the first letter of the entire title capitalized}
%Subsections only have the first letter of the first word capitalized
%(except for words that are naturally capitalized).

\section{Galactic Centre pulsar searches}
Over the last couple of decades a number of searches of the inner tens
of parsecs have taken
place\cite{jkl+06,dcl+09,mkf+10,sbb+13,ekk+13}. In 2013, the Galactic
Centre (GC) pulsar population\footnote{Here we define the GC pulsar
  population as those with pulses detected in radio at a projected
  offset of less than 0.5$^\circ$ ($\sim$73$\,{\rm pc}$) from Sgr~A*.}
was brought to a total of six with the detection of PSR~J1745$-$2900,
first through its gamma ray emission\cite{kbk+13,mgz+13} and then
finally in radio\cite{efk+13,sj+13}.

As radio telescope hardware and analysis techniques are improved, GC
pulsar searches are carving out additional areas of parameter
space neglected by previous searches. An example is the progression to
higher than ``normal'' pulsar observing frequencies ($f_{\rm
  obs}\gtrsim10\,$GHz) where the deleterious effects of interstellar
dispersion and scattering (effects that are at their highest in GC)
are reduced. Unfortunately, improvements in sensitivity by
choosing higher frequencies are limited by the steep
spectrum of pulsar emission, where detection signal-to-noise ratio
${\rm S/N}$ is typically $\propto f_{\rm obs}^{\sim -1.7}$. This is one
case where competing selection effects affect the performance of
GC pulsar searches (Table~\ref{tab1}).  In general, higher
observing frequencies and longer integration lengths $T_{\rm obs}$
(given the large distance of the GC) are more favourable, however this
predisposes searches to flatter spectrum isolated pulsars. Binary
pulsar searches can become computationally prohibitive above certain
$T_{\rm obs}$ where even single parameter acceleration search
computations $C_{a}$ scale with $T_{\rm obs}^3$.

\begin{table}
\tbl{Competing observational selection effects for GC radio pulsar
  searches. Pulse broadening effects, such as dispersion smearing
  $\tau_{\rm ch}$, scatter broadening $\tau_{\rm s}$ and acceleration broadening $\tau_{a}$, add in
  quadrature to the intrinsic pulse width $W$ making $W_{\rm eff}$,
  which scales with the detection signal-to-noise ratio as ${\rm S/N}
  \propto \sqrt{(P-W_{\rm eff})/W_{\rm eff}}$ where $P$ is the spin
  period. \cmark \hspace{1pt} indicates favourable observational
  configurations while \xmark \hspace{1pt} are unfavourable.}
{\begin{tabular}{@{}lcc@{}}
\toprule
\multicolumn{1}{c}{Observing frequency related} & Higher $f_{\rm obs}$ & Lower $f_{\rm obs}$\\
\colrule
Intra-channel pulse dispersion smearing ($\tau_{\rm ch} \propto f_{\rm obs}^{-3}$) & \cmark & \xmark \\
Pulse scattering broadening  ($\tau_{\rm s} \propto f_{\rm obs}^{-3.8}$) & \cmark & \xmark \\
Intense GC background continuum emission\cite{law+08} (${\rm S/N} \propto f_{\rm obs}^{-0.8}$) & \cmark & \xmark \\
Pulsar spectra (${\rm S/N} \propto f_{\rm obs}^{\sim -1.7}$) & \xmark & \cmark \\
\colrule
\multicolumn{1}{c}{Integration length related} & Longer $T_{\rm obs}$ & Shorter $T_{\rm obs}$\\
\colrule
GC distance (${\rm S/N} \propto \sqrt{T_{\rm obs}}$) & \cmark  & \xmark \\
Pulse broadening from acceleration ($\tau_{a} \propto T_{\rm obs}^2$) & \xmark & \cmark \\
Pulse broadening from jerk ($\tau_{j} \propto T_{\rm obs}^3$) & \xmark & \cmark \\
Computational operations for acceleration search ($C_{a} \propto T_{\rm obs}^3$)& \xmark & \cmark \\
\botrule
\end{tabular}}
%\begin{tabnote}
%$^{\text a}$ Sample table footnote.\\
%\end{tabnote}
\label{tab1}
\end{table}

\section{BlackHoleCam pulsar work}
BlackHoleCam\cite{goddi+17}, and members of the Event Horizon Telescope (EHT)
Consortium, aim to tackle these observational constraints by using the
largest and most sensitive telescopes operating at millimetre
wavelengths which form elements of the EHT telescope. 
%PSR~J1745$-$2900 has demonstrated that pulsars in the GC can be
%observed at millimetre wavelengths.
Because both EHT VLBI imaging and pulsar observations can utilise the
same raw data product from each array element, EHT VLBI and pulsar
observations can be commensal.

Thus far, pulsar search efforts have concentrated on analysing data
from the single most sensitive element of the EHT: ``fully phased''
ALMA. In the future we can envisage using a phased array of the
largest components of the EHT to further increase sensitivity or to
mitigate site specific interference contamination. 

\section*{Acknowledgments}
The authors acknowledge financial support by the European Research
Council (ERC) Synergy Grant ``BlackHoleCam: Imaging the Event Horizon
of Black Holes'' (grant 610058).

\end{document}